\RequirePackage{fix-cm}

\documentclass[smallextended]{svjour3}       % onecolumn (second format)
\smartqed 
\usepackage{graphicx,hyperref,amsmath}

\begin{document}

\title{{O}n {C}hallenges to {S}eparability of the {D}irac {E}quation in {K}err {G}eometry under {C}ompact {H}yperboloidal {C}oordinates}

\author{Aditya Tamar}

\institute{Independent Researcher \at
              124 Bank Enclave, First Floor, Laxmi Nagar, Delhi-110092, India \\
              \email{adityatamar@gmail.com} \\
              ORCID ID: 0000-0001-8763-4169
}

\date{Received: date / Accepted: date}

\maketitle

\begin{abstract}
The Dirac equation governs the behaviour of spin-1/2 particles. The equation's separability into decoupled radial and angular differential equations is a crucial step in analytical and numerical computations of quantities like eigenvalues, quasinormal modes and bound states. However, this separation has been performed in co-ordinate systems that are not well-behaved in either limiting regions of $r \rightarrow r_{horizon}$, $r \rightarrow r_\infty$ or both. In particular, the extensively used Boyer-Lindquist co-ordinates contains unphysical features of spacetime geometry for both   $r_{horizon}$ and $r_\infty$. Therefore, motivated by the recently developed compact hyperboloidal co-ordinate system for Kerr Black Holes that is well behaved in these limiting regions, we attempt the separation of the Dirac equation. We first construct a null tetrad suitable for the separability analysis under the Newman-Penrose formalism.  Then, an unexpected result is shown that by using the standard separability procedure based on the mode ansatz under this tetrad, the Dirac equation does not decouple into radial and angular equationsPossible reasons for this behaviour as well as importance of proving separability for various computations are discussed.
\keywords{Separation of Dirac equation in curved space \and Kerr Black Hole \and Hyperboloidal Coordinates \and Newman-Penrose formalism}
\end{abstract}
\newpage
\section{Introduction}
\label{intro}
The Dirac equation \cite{Dirac1928} is an equation governing the nature of  spin-1/2 particles (fermions). While it is almost certain that the equation describes charged fermions, deciding whether uncharged fermions (i.e neutrinos) are Dirac or Majorana particles is a field of active research in particle physics \cite{Neutrino2020}, quantum information sciences\cite{QIMajorana} and astrophysics \cite{NeutrinoAstro}. In a similar vein, the Dirac equation has been used in different branches of physics, wherein its analysis in the presence of a strong gravitational fields has provided insight into various phenomenon of mathematical and physical interest \cite{Finster1,Finster2,Finster3}. One of the most exciting areas of study is the behaviour of fermions around Black Holes\cite{Finster1,Finster2,Finster3,Dolan2015,Roeken2017,Unruh1973,Chandra1976,Page1976}, and a crucial step in this analysis is the separation of the Dirac equation into decoupled radial and angular Ordinary Differential Equations (ODE). This separation allows us to investigate various qualitative features such as bound states \cite{Dolan2015}, decay rates \cite{Finster3} and quasinormal modes \cite{Konoplya2017,Coutant2019,Dogan2019}. It has also been used for providing insight into non-trivial extensions of special mathematical functions \cite{Kraniotis2019}. Now this crucial separation of equations was first performed by Unruh in 1973 using the 4-spinor formalism \cite{Unruh1973} for the massless field in the Kerr metric\cite{Kerr1963}. This was followed by Chandrashekhar \cite{Chandra1976} and Page \cite{Page1976} for the massive case, both using the 2-spinor or Newman Penrose formalism \cite{NewPen1962} for Kerr and Kerr-Newman metric \cite{KerrNewman} respectively. It is to be noted that the separation of variables under perturbations cannot be conducted for arbitrary metric. For this to happen, the metric should possess certain symmetries that are manifest by the presence of certain Killing vectors \cite{Carter1968} and Killing-Yano tensors \cite{Sommers}.

Now, while the aforementioned analysis has provided a rich theoretical framework, their regime of validity is restricted by the nature of the co-ordinates used. With very few exceptions \cite{Dolan2015,Roeken2017} amongst the limited studies of the Dirac equation around Kerr Black Holes, the Kerr metric is described using the Boyer-Lindquist (BL) co-ordinates $ \{t,r,\theta,\phi \} $ \cite{BoyLind}. In fact, the famous separation by Chandrashekhar in \cite{Chandra1976}  was performed precisely using these co-ordinates. The primary reason for using the BL co-ordinates is the relative simplicity of the expressions used in analytical calculations. However, the BL co-ordinates are not well behaved, both for $r \rightarrow r_\infty$ and $r \rightarrow r_{horizon}$.

Therefore, motivated by the advances in the numerical relativity for dealing with these boundary conditions, \cite{Hilditch1,Hilditch2,Macedo2016,Zenginoglu2008,Macedo2018,Bieri2020,Ruchlin2017}, we study the separation of the Dirac equation in Kerr Geometry that is subjected to a hyperboloidal formulation \textit{and} gauge fixing using the height function technique \cite{Zenginoglu2008}. This establishes a co-ordinate system that is well-behaved for both, through the horizon (i.e "horizon penetrating") as well as at null infinity, $\mathcal{J^+}$. These regions are extremely crucial in establishing the boundary conditions for quasinormal modes \cite{QNMBound,Konoplya2018,Dolan2015,QNMBound2} and therefore it is expected that further advancements using this well-defined co-ordinate system will provide new insights in the numerical and analytical treatment of black hole perturbations. 

This paper is organised as follows. In Sec. \hyperlink{section.2}{II} we describe the Kerr metric in Boyer-Lindquist, Ingoing-Kerr co-ordinates and discuss their qualitative features and shortcomings. In Sec. \hyperlink{section.3}{III} we describe the physical motivations behind constructing a compact hyperboloidal framework and then construct a Newman-Penrose tetrad in this co-ordinate system. We explain how this addresses the shortcomings described in Sec. \hyperlink{section.2}{II}. In Sec. \hyperlink{section.4}{IV} we describe the Dirac equation in Kerr Geometry under the Newman Penrose formalism and attempt its separation using the tetrad of Sec. \hyperlink{section.3}{III}. It is shown that the Dirac equation does not separate under the standard separation procedure. In Sec. \hyperlink{section.5}{V} we discuss the possible reasons for this behaviour along with the need of demonstrating separability for future work using compact hyperboloidal co-ordinates.

 \section{Kerr Metric}  
This section discusses the Kerr Metric in Boyer-Linqduist (BL) and Ingoing Kerr, which is also known as Advanced Eddington-Finkelstein co-ordinate system. We utilise the terminology of \cite{Macedo2016} for all co-ordinate systems.
\subsection{In {B}oyer-{L}indquist {C}o-ordinates}
The Kerr Metric in BL co-ordinates \{$t,r,\theta,\varphi$\} is given in most standard literature in the following form:
\begin{gather}
ds^2 = -fdt^2 - \frac{4Mar}{\Sigma}\sin^2\theta dtd\varphi + \frac{\Sigma}{\Delta}dr^2 + \Sigma^2d\theta^2 \nonumber \\ 
+\sin^2\theta \bigg(\Sigma_0 + \frac{2Ma^2r}{\Sigma}\sin^2\theta \bigg)d\varphi^2 \label{eq1}
\end{gather}
where, 
\begin{gather}
\Delta(r) = r^2 - 2Mr + a^2, \Sigma(r,\theta)= r^2 +a^2\cos^2\theta, \nonumber \\ 
\Sigma_0(r) = r^2 +a^2, f(r,\theta) = 1-\frac{2Mr}{\Sigma(r,\theta)}  \label{eq2}
\end{gather}
Here, $M$ is the mass and $a$ is the angular momentum. of the black hole The roots of the equation $\Delta(r)=0$ correspond to the event horizon $r_+$ and Cauchy horizon $r_-$. By substituting $a=0$, we obtain the non-rotating, spherically symmetric Schwarzschild metric.\\
As noted earlier, along the $t$=constant hypersurface, $r \rightarrow \infty$ leads to spatial infinity $i^0$ which contains geometric features of both future null-infinity$\mathcal{J^+}$and past null infinity $\mathcal{J^-}$ whereas $r \rightarrow r_+$ leads to the bifurfaction sphere $\mathcal{B}$ of the black hole that contains geometric features of both black hole $\mathcal{H^+}$ and "white hole" horizon $ \mathcal{H^-}$. However, astrophysical black holes do not have a bifurcation sphere and observers of gravitational radiation in the far away region do not have access to spatial infinity \cite{Zenginog2011}. Also from a purely algebraic standpoint, the metric is singular at both $r_+$ and $r_-$ (since they are roots of $\Delta (r) = 0$) and therefore unsuitable for studying the behaviour at the horizon.
For an extended discussion on the properties of Kerr Black Holes in BL co-ordinates we refer the reader to \cite{BHPhysics}.

\subsection{In ingoing {K}err {C}o-ordinates}
In order to partially rectify the shortcomings of the BL co-ordinates, we can introduce co-ordinates that are well-behaved at the black hole horizon. These co-ordinates are called Ingoing-Kerr \cite{Macedo2016}, Advanced Eddington-Finkelstein co-ordinates\cite{Roeken2017} or Ingoing Eddington-Finkelstein co-ordinates \cite{Macedo2018,MWT}. An excellent description of the features of these co-ordinate systems is given in \cite{MWT}. To transform to ingoing-Kerr, we perform the following transformation to $t$ and $\phi$:  
\begin{gather}
t = v -r^*(r), \varphi = \phi - k(r) \label{eq3}
\end{gather}
where the tortoise coordinate $r^*(r)$ and phase factor $k(r)$ are defined by,
\begin{gather}
\frac{dr^*}{dr} = \frac{\Sigma_0}{\Delta},  \frac{dk}{dr} = \frac{a}{\Delta}
\end{gather}
The line element is given by:
\begin{gather}
ds^2 = -f \bigg(dv - a\sin^2\theta d\phi \bigg)^2 + \Sigma d\omega^2 \nonumber \\
+2 \bigg(dv - a\sin^2\theta d\phi \bigg) \bigg(dr - a\sin^2\theta d\phi \bigg)
\end{gather}
where, 
\begin{gather}
d\omega^2 = d\theta^2 + \sin^2\theta d\phi
\end{gather}
In terms of spacetime geometry, it fixes the behaviour on the horizon; the hypersurface $r=r_+$ along $v=$ constant corresponds \textit{only} to $\mathcal{H^+}$. However, the limiting region of $r \rightarrow \infty$ leads to $\mathcal{J^-}$ wherein for the physical case it should correspond to $\mathcal{J^+}$. 

Before we fix this pathology in the next section, we shall introduce a set of null vectors ${l,n,m, \bar{m}}$ where $l$ and $n$ correspond to outgoing and ingoing null vectors respectively and $m,\bar{m}$ are complex conjugates of each other. We will be utilising this co-ordinate system as our basis for hyperboloidal compactification in the next section. We follow the construction of \cite{Macedo2020,KrishnanCoord} suited to our notation. The contravariant components are: 

\begin{gather}
l^a = \frac{1}{\zeta} \bigg(\frac{r^2 + a^2}{\Sigma}\frac{\partial}{\partial v} + \frac{\Delta}{2\Sigma}\frac{\partial}{\partial r} + \frac{a}{\Sigma}\frac{\partial}{\partial \phi} \bigg) \\
n^a = - \zeta \frac{\partial}{\partial r}\\
m^a = \frac{a\sin \theta}{\sqrt{2} \Sigma^+}\frac{\partial}{\partial v} + \frac{i}{\sqrt{2}\Sigma^+}\frac{\partial}{\partial \theta} + \frac{\Sigma^+ \sin \theta}{\sqrt{2} }\frac{\partial}{\partial \phi}\\
\bar{m}^a = \frac{a\sin \theta}{\sqrt{2} \Sigma^-}\frac{\partial}{\partial v} - \frac{i}{\sqrt{2}\Sigma^-}\frac{\partial}{\partial \theta} + \frac{\Sigma^- \sin \theta}{\sqrt{2}} \frac{\partial}{\partial \phi}
\end{gather}
where,
\begin{gather}
\Sigma^+ = r +  ia\cos \theta, \Sigma^- = r - ia\cos \theta 
\end{gather}
Here $\zeta$ is a boost parameter that will be fixed in the next section. It can be verified that the null tetrad satisfies the normalisation conditions $l.n = -1$ and $m.\bar{m} = 1$ (some treatments in literature follow $l.n = 1$ and $m.\bar{m} = -1$) with all other inner products being $0$. Satisfying these conditions is necessary for the null tetrad to be a Newman Penrose tetrad which forms the basis of our separation analysis. 
\section{Compact Hyperboloidal Co-ordinates}
\subsection{Hyperboloidal Foliation}
The basic idea of hyperboloidal foliation is as follows \cite{Zenginoglu2008}. For an asymptotically flat spacetime of manifold $\mathcal{\tilde{M}}$ and metric $\tilde{g}$, we can attach a boundary condition at null infinity $\mathcal{J}$ (i.e both future and past null infinity) such that a rescaling of $\tilde{g}$ by a conformal factor $\Omega$ would result in a metric $g= \Omega^2 \tilde{g}$ that behaves smoothly through $\mathcal{J}$. The conformal factor shall be positive definite throughout the manifold.  The triplet $(\mathcal{M},g,\Omega)$ where $\mathcal{M}$ is the manifold endowed with the boundary $\Omega$ is known as the \textit{conformal extension} of $(\mathcal{\tilde{M}}, \tilde{g})$ and the latter is now called a weakly asymptotically simple spacetime. All physically reasonable solutions of the Einstein field equations are weakly asymptotically simple. 

Now, to perform the foliation at null-infinity we introduce spacelike surfaces that can extend through $\mathcal{J^+}$. They are called hyperboloidal surfaces as they asymptotically behave as hyperboloids in Minkowski spacetime \cite{Friedrich1983}. In order to use the hyperboloidal foliation for studying radiation phenomenon of fields, we utilise a suitable gauge in which the spatial co-ordinate location of $\mathcal{J}$ is independent of time. Therefore, our time function transforms under a \textit{height function} $h(r)$. These foliations have an extremely rich mathematical structure and we refer to \cite{Zenginoglu2008, Zenginog2011} for details of their mathematical structure as well as computational advantages.

\subsection{Co-ordinate System for Kerr Black Holes}
Keeping in mind the discussion in Sec. \hyperlink{subsection.3.1}{3.1}, we obtain compact hyperboloidal co-ordinates $ {\tau,\sigma,\theta,\varphi} $ perform the following transformations: 
\begin{gather}
v = \lambda \bigg( \tau - h(\sigma, \theta) \bigg), r = \lambda \frac{\rho(\sigma)}{\sigma},
\end{gather}
where $\lambda$ has the length scale of spacetime, $h(\sigma,\theta)$ is our height function and $\rho$ is the radial gauge freedom. In terms of the new radial co-ordinate $\sigma$, we introduce a conformal factor $\Omega$, 
\begin{gather}
\Omega = \frac{\sigma}{\lambda}
\end{gather}
This rescales the line element as $\tilde{ds}^2 = \Omega^2 ds^2$:
\begin{gather}
\tilde{ds}^2= -\sigma^2F \bigg( d\tau - h_,\sigma d\sigma - h_,\theta d\theta -\alpha \sin^2 \theta d\phi \bigg)^2 + \tilde{\Sigma}^2 d\omega^2 \nonumber \\
-2 \bigg( d\tau - h_,\sigma d\sigma - h_,\theta d\theta - \alpha \sin^2 \theta \bigg) \bigg( \beta d\sigma + \alpha \sigma^2 \sin^2 \theta d\phi \bigg)
\end{gather}
where,  $d\omega^2 = d\theta^2 + \sin^2\theta d\phi^2$ and $\beta = \rho(\sigma) - \sigma \rho'(\sigma)$. Under conformal scaling, the notation in \eqref{eq2} transform as follows: 
\begin{gather}
\tilde{\Delta}(\sigma) = \Omega^2 \Delta(r(\sigma)) = \rho (\sigma)^2 -2\mu \rho(\sigma) + \alpha^2\rho(\sigma)^2,\\
\tilde{\Sigma}(\sigma,\theta) = \Omega^2\Sigma(r(\sigma),\theta) = \rho(\sigma)^2 + \alpha^2\sigma^2\cos^2\theta,\\
\tilde{\Sigma_0} = \Omega^2\Sigma_0(r(\sigma)) = \rho(\sigma)^2 + \alpha^2\sigma^2.\\
F(\sigma, \theta) = f(r(\sigma), \theta) = 1-\frac{2\mu\rho(\sigma)\sigma}{\tilde{\Sigma}(\sigma,\theta)}
\end{gather}
where $\mu =\frac{M}{\lambda}$ and $\alpha = \frac{a}{\lambda}$ are dimensionless mass and spin parameters respectively.
In this paper, we operate in the radial function fixing minimum gauge, which assigns values to the various parameters introduced in the above metric. A description of the minimum gauge,  the corresponding metric components and the values of the parameters in the radial gauge are provided in \cite{Macedo2020}. The co-ordinate system in this gauge transforms to the following null-vectors:
\begin{gather}
\tilde{l}^a = \bigg(\frac{h_R(2\tilde{\Sigma_0} -\tilde{\Delta} \sigma^2 h_R)}{2\tilde{\Sigma}}\bigg) \frac{\partial}{\partial \tau} - \frac{\tilde{\Delta}\sigma^2 h_R}{2\tilde{\Sigma}}\frac{\partial}{\partial \sigma} + \frac{\alpha \sigma^2 h_R}{\tilde{\Sigma}}\frac{\partial}{\partial \phi} \\
\tilde{n}^a = \frac{\partial}{\partial \tau} + \frac{1}{h_R}\frac{\partial}{\partial \sigma}\\
\tilde{m}^a = \frac{a\sin \theta}{\sqrt{2} \Sigma^+}\frac{\partial}{\partial v} + \frac{i}{\sqrt{2}\Sigma^+}\frac{\partial}{\partial \theta} + \frac{\Sigma^+ \sin \theta}{\sqrt{2} }\frac{\partial}{\partial \phi}\\
\tilde{\bar{m}}^a = \frac{a\sin \theta}{\sqrt{2} \Sigma^-}\frac{\partial}{\partial v} - \frac{i}{\sqrt{2}\Sigma^-}\frac{\partial}{\partial \theta} + \frac{\Sigma^- \sin \theta}{\sqrt{2}} \frac{\partial}{\partial \phi}
\end{gather}
where, $h_R$ is the height function in the radial function fixing gauge. The parameter $\zeta$ was fixed by the conditions described in \cite{Macedo2020}. From the covariant metric components $\tilde{g}_{ab}$ for the minimum gauge we obtain the following covariant vectors:
\begin{gather}
\tilde{l}_a = -\frac{C(\sigma)}{2\tilde{\Sigma}}d\tau + \bigg( \frac{h_R(C(\sigma)-2\Sigma)}{2\tilde{\Sigma}} \bigg)d\sigma  + \frac{C(\sigma)\alpha\sin^2\theta}{\tilde{2\Sigma}}d\phi \\
\tilde{n}_a = -\bigg( \frac{1}{h_R} \bigg)d\tau + d\sigma + \bigg(\frac{\alpha \sin^2 \theta}{h_R} \bigg)d\phi \\
\tilde{m}_a = -\frac{\alpha \sigma^2\sin\theta}{\sqrt{2} \tilde{\Sigma}^+}d\tau + \frac{\alpha \sigma^2\sin\theta h_R}{\sqrt{2}\tilde{\Sigma}^+}d\sigma+ \frac{i\tilde{\Sigma}}{\sqrt{2}\tilde{\Sigma}^+}d\theta \nonumber \\ + \frac{\tilde{\Sigma_0}\sin\theta}{\sqrt{2}\tilde{\Sigma}^+}d\phi \\
\tilde{\bar{m}}_a = -\frac{\alpha \sigma^2\sin\theta}{\sqrt{2} \tilde{\Sigma}^-}d\tau + \frac{\alpha \sigma^2\sin\theta h_R}{\sqrt{2}\tilde{\Sigma}^-}d\sigma - \frac{i\tilde{\Sigma}}{\sqrt{2}\tilde{\Sigma}^-}d\theta \nonumber \\ + \frac{\tilde{\Sigma_0}\sin\theta}{\sqrt{2}\tilde{\Sigma}^-}d\phi
\end{gather}
where, in order to avoid an overabundance of Greek symbols at a later stage we use the notation:
\begin{gather}
C(\sigma) = \tilde{\Delta}\sigma^2 h_R
\end{gather}
It can be checked that the null vectors still satisfy the normalisation conditions $\tilde{l}.\tilde{n} = -1$ and $\tilde{m}.\tilde{\bar{m}} = 1$ with all other inner products being zero.
\section {Dirac Equation in Newman-Penrose Formalism}
The Dirac equation in the Newman-Penrose formalism is given by \cite{ChandraBH}: 
\begin{gather}
(D + \varepsilon - \rho)F_1 + (\delta^* + \pi - \alpha)F_2 = i\mu_* G_1;\\
(\Delta^0 + \mu - \gamma)F_2 + (\delta + \beta - \tau)F_1 = i\mu_*G_2;\\
(D + \varepsilon^* - \rho^*)G_2 - (\delta + \pi^* - \alpha^*)G_1 = i\mu_*F_2;\\
(\Delta^0 + \mu^* - \gamma^*)G_1 - (\delta^* + \beta^* - \tau^*)G_2 = i\mu_*F_1;
\end{gather}
Here, $D,\Delta^0$ are purely radial differential operators for $\tilde{l}$ and $\tilde{n}$ respectively, whereas $\delta$ and $\delta^0$ are purely angular differential operators for $\tilde{m}$ and $\tilde{\bar{m}}$ respectively. The quantities with $*$ denote the complex conjugate. $F_1, F_2,G_1,G_2$ correspond to the components of the Dirac wavefunction in 2-spinor formalism.  The factor $\mu_*\sqrt{2}$ corresponds to the mass of the spin-1/2 particle expressed in inverse Compton wavelength. The remaining Greek symbols correspond to the spin coefficient for a given metric. A description of their physical relevance is given in \cite{PenroseRindler}. While there are several ways to calculate them based on differential geometry\cite{ChandraBH,PenroseRindler}, we found is most straightforward to use the method in \cite{Cocke1989} which is solely based on partial derivatives of the null-vectors. The spin coefficients for our co-ordinates are as follows: 
\begin{gather}
\kappa = \nu = \sigma = \lambda = 0;\\
\varepsilon = \frac{\tilde{\Sigma}C'(\sigma)-2C(\sigma)\alpha^2\sigma \cos^2\theta}{4\tilde{\Sigma}^2} + \frac{1}{2} \rho \\
\pi = \frac{\alpha \sigma \sin \theta}{\sqrt{2}\tilde{\Sigma}}\\
\beta = \frac{2\alpha\sigma\sin\theta - i\cot\theta\tilde{\Sigma}}{2\sqrt{2}\tilde{\Sigma}\tilde{\Sigma}^-},\\
\alpha = \beta^*,\\
\mu = -\frac{i\alpha\cos\theta}{h_R\tilde{\Sigma}^+},\\
\tau = \pi,
\gamma = \frac{1}{2}\mu,\\
\rho = \frac{C(\sigma)\alpha\cos\theta[\alpha\sigma\cos\theta(1+\tilde{\Sigma})-2i]}{4\tilde{\Sigma^2}}
\end{gather}
\section{Challenges to Separability and Conclusion}
 When the author attempted the separation of the Dirac equation based on the standard mode ansatz of  \cite{ChandraBH}, there seem to be two technical challenges. Firstly, the terms $ \varepsilon - \rho$ has a rather complex form with a seemingly coupled relationship between radial and angular variables. Secondly, assuming that this coupling can be resolved and we perform the mode analysis, it is rather unclear how the radial and angular variables would be separated in the later stages since due to conformal scaling the factors $\tilde{\Sigma^+}, \tilde{\Sigma^-}$ and $\tilde{\Sigma}$ have an inherent coupling in the variables $\sigma, \theta$. This is in stark contrast to the nature of equivalent factors in the BL and Ingoing Kerr co-ordinates in which the radial and angular variables remain decoupled from each other. Therefore, even after resolving the first technical difficulty, separation remains suspect. 
 
Thus to conclude, in this short article we have demonstrated a rather unexpected result that the separability of the Dirac equation seems to be heavily dependent on the co-ordinate system being employed. In a physically well-behaved co-ordinate system i.e the compact hyperboloidal co-ordinates, the Dirac equation has shown to not admit separability.This result is to the best of the author's knowledge the only existing case in literature in which for a given co-ordinate system the Dirac equation has not been separable into radial and angular equations.

Furthermore, the separability of the Dirac equation is a crucial step in the analysis of massive and massless spin-1/2 particles in curved space. The resolution of the aforementioned technical difficulties is extremely crucial for further analytic and numerical computations of quantities such as quasinormal modes and eigenvalues using this co-ordinate system. 

\begin{acknowledgements}
I would like to thank Vedavathi P. for her continued support.
\end{acknowledgements}

\bibliographystyle{spmpsci}

\end{document}